\documentclass[%
 reprint,
 amsmath,amssymb,
 aps,
pre,
]{revtex4-2}

\usepackage{graphicx}
\usepackage{dcolumn}
\usepackage{bm}
\usepackage{hyperref}

\usepackage[english]{babel}
\usepackage{blindtext}
\usepackage{xcolor}

\begin{document}

\title{Accelerating Multicanonical Sampling with Irreversibility}

\author{Thomas Vogel}
\email{E-mail: thomasvogel@lanl.gov}
\thanks{Current address: Computer, Computational, and Statistical Sciences Division, Los Alamos National Laboratory, Los Alamos, NM 87545, U.S.A.}
\affiliation{%
Department of Physics and Astronomy, University of North Georgia, Dahlonega, GA 30597, U.S.A.
}%

\author{Ying Wai Li}
\email{E-mail: yingwaili@lanl.gov}
\affiliation{%
Computer, Computational, and Statistical Sciences Division, Los Alamos National Laboratory, Los Alamos, NM 87545, U.S.A.
}%

\date{\today}

\begin{abstract}
Flat-histogram Monte Carlo simulations are well-established, robust methods to perform random walks in a physical observable or parameter space, making them suitable for finding ground states or studying phase transitions in complex systems in statistical physics. However, their efficiency can be limited by the time to attain the desired flat distribution, which is generally unknown prior to the simulations. In particular, they might suffer from slowing down towards the end of a~simulation due to the diffusive nature of random walks. In this work we apply irreversibility to the multicanonical Monte Carlo method via the lifting approach to alleviate this behavior. We achieve a~2--4 times speedup in ground-state search for a two-dimensional (2D) Ising model, and up to an order of magnitude of speedup for finding the ground-state energy in an Edwards--Anderson spin glass, compared to traditional multicanonical sampling. The round-trip times between ground states show a narrower distribution and are significantly shorter compared to the reversible counterpart, suggesting that a lower convergence time with a smaller time variance is feasible.
\end{abstract}

\maketitle



\section{Introduction}

Markov-Chain Monte Carlo (MCMC) algorithms statistically sample from a target distribution by sequentially proposing updates to the current state of a system, accepting or rejecting each proposal, or Monte Carlo (MC) trial move, according to the probability distribution.
Long autocorrelation times are the root cause of the notorious inefficiency of many MCMC methods, particularly when studying physical systems near phase transitions~\cite{wolff90npb,janke1998pa,janke08lnp}.

\begin{figure}[b!]
    \centering
    \includegraphics[width=\columnwidth]{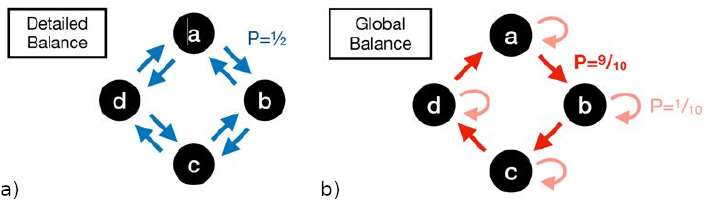}
    \caption{a) Transitions between states are equally likely in both directions (detailed balance), a walk through those states is always ``reversible''. b) Transitions between states are only possible in one directions, yet the probability ``influx'' equals the ``outflux'' in each state (global balance).}
    \label{fig:DB-GB}\vskip-0.5mm
\end{figure}

At equilibrium, let $P_a$ denote the stationary probability of the system being in state~$a$, and let $W_{a\rightarrow b}$ denote the transition probability to move from state $a$ to $b$ in a MC trial move, then we can ensure the conservation of probability by ensuring ``detailed balance'' between the two states $a$ and $b$, that is,
\begin{equation}
P_a\, W_{a\rightarrow b}=P_b\, W_{b\rightarrow a}
\label{eq:db}
\end{equation}
for every given trial move. In particular, there must always be a non-zero probability to immediately go back to the previous state for every MC move (``reversibility''). In traditional MC algorithms such as Metropolis--Hastings~\cite{metropolisEquationStateCalculations1953,hastingsMonteCarloSampling1970}, the detailed balance condition (Eq.~\ref{eq:db}) holds: the probability flux from state $a$ to state $b$ always exactly matches the reverse one (see also Fig.~\ref{fig:DB-GB}\,a). 

While detailed balance is sufficient to sample $P_a$ for all states $a$, it is an overly constraining condition.
A less stringent scheme is for the transition probability to fulfill ``global balance'':
\begin{equation}
\sum_{b\neq a} P_a\, W_{a\rightarrow b}=\sum_{b\neq a} P_b\, W_{b\rightarrow a}\,.
\label{eq:gb}
\end{equation}
This leaves the distributions $P_a$ and $P_b$ invariant, without necessarily obeying detailed balance (see, for example, Ref.~\cite{landau2014book} for a more thorough introduction of the theory behind MCMC sampling and~\cite{turitsynIrreversibleMonteCarlo2011,weigel2010cpc,krauth14jcp}, and below, for MC algorithms employing this more general balance condition). Markov chains that satisfy only global balance and allow for directional flux between any state-pair ($a,b$) are called \textit{irreversible} (Fig.~\ref{fig:DB-GB}\,b). 

Removing the restrictions of detailed balance allows for much more flexibility in the design of MC move proposals and acceptance criteria, opening a wide range of possibilities for more efficient algorithm design. One strategy to impose irreversibility is the concept of ``lifting'', a method that reduces the diffusive nature of random walks~\cite{diaconisAnalysisNonreversibleMarkov2000, faiziEfficientIrreversibleMonte2020, chenLiftingMarkovChains1999, turitsynIrreversibleMonteCarlo2011, vuceljaLiftingNonreversibleMarkov2016}. 
Lifting duplicates the phase space to create two flows, each biased in opposite directions. 
They were successfully used in combination with Metropolis MC sampling for discrete distributions such as Ising and Potts models~\cite{weigel2010cpc,faiziEfficientIrreversibleMonte2020}.
The class of event-chain Monte Carlo algorithms further extends lifting to continuous-time and continuous-space systems such as hard-spheres, dense soft-matter systems, spin models, and systems with arbitrary pairwise interactions~\cite{bernard_krauth_pre_2009, with12pre, krauth14jcp, nishikawa15pre, harland17epl, kampmann21fp, krauth21fp, maggs_krauth_pre_2022}.
In this work we apply the lifting strategy, adapted earlier in Ref.~\cite{weigel2010cpc} for canonical ensembles, in a multi\-canonical setting to the 2D Ising and Edwards--Anderson spin glass models. We show that the new algorithm delivers correct results and leads to a speedup of up to a factor of ten in Monte Carlo time for finding ground states and exploring the phase space of these spin systems.

\section{\label{sec:method}Lifting in Multicanonical Sampling}

\subsection{Multicanonical sampling}

In canonical ensembles, the statistical weights of states are determined by the Boltzmann factor. The canonical distribution in energy reads
\begin{equation}
    P_{\textrm{can}}(E)\propto g(E)\,e^{-\beta E}\,,
\end{equation}
where $\beta=1/k_BT$, with $k_B$ being the Boltzmann constant, $T$ being the temperature, $E$ is the total energy of the system according to the Hamiltonian, and $g(E)$ is the density of states in energy. In multicanonical simulations~\cite{berg1991plb,berg1992prl,janke1998pa}, as in other ``flat histogram'' methods~\cite{wang2001prl:wl,zhou2006prl:wl,vogel2013prl,junghans2014jctc}, one aims at replacing the Boltzmann factor by sampling weights $W(E)$ that are determined such that the distribution of energy states, $P_{\textrm{muca}}(E)$, is, ideally, flat:
\begin{equation}
    P_{\textrm{muca}}(E)\propto g(E)W(E)=\textrm{const}.
    \label{eqn:muca}
\end{equation}
In principle, this can be achieved by setting $W(E)=1/g(E)$. 
However, in practice, $g(E)$ is typically not known \textit{a priori}. The weights $W(E)$ can then be estimated in an iterative way. One way is to initialize the weights $W_0(E)$ from a canonical distribution. In each iteration $i$, one samples from the distribution to accumulate a histogram $H_i(E)$. MC moves are accepted with probability:
\begin{equation}
    p=\min\left[1,\frac{W_i(E_{\textrm{after}})}{W_i(E_{\textrm{before}})}\right]\,.
\end{equation}
where $E_\textrm{before/after}$ denotes the total energy of the system before and after a MC move is performed. At the end of each iteration, one updates the weights as $W_{i+1}(E) = W_i(E)/H_i(E)$ to broaden the sampled energy range in the next iteration~\cite{janke1998pa,zierenberg2013cpc}. At the end of such a scheme, the density of states $g(E)$ can be estimated from collecting a histogram in a typically longer ``production'' run without updating the weights anymore and applying Eq.~(\ref{eqn:muca}). This is the method we employed in this work. Alternatively, one could run one (or a few) iterations of the Wang--Landau algorithm~\cite{wang2001prl:wl,wang2001pre:wl} to obtain an initial estimate of the weights for a~subsequent multicanonical simulation.

\subsection{A procedure for lifting}

\begin{figure*}
    \centering
    \includegraphics[width=0.72\textwidth]{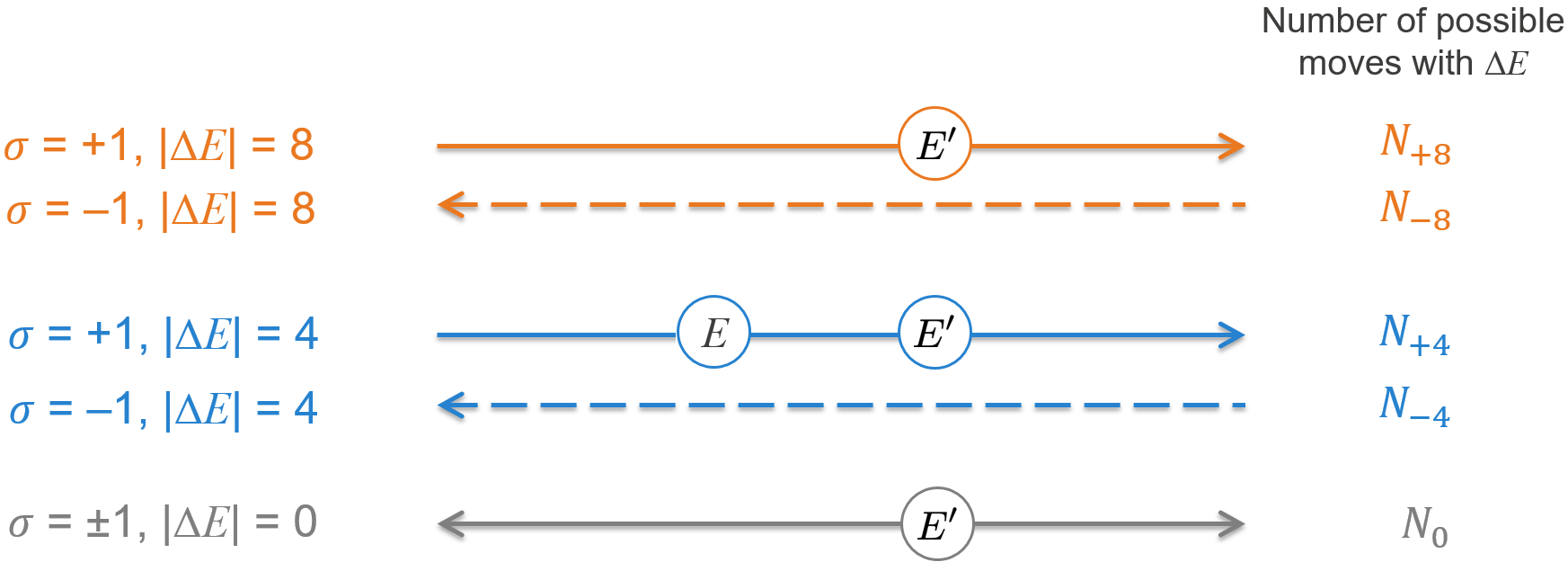}
    \caption{The lifting scheme for a 2D Ising model consisting of separate chains corresponding to changes in energy $|\Delta E|=0,4,$ and~$8$. Each $|\Delta E| \neq 0$ chain has a direction, $\sigma = +1$ or $\sigma= -1$, where a MC move results in an energy change; whereas on the $|\Delta E|=0$ chain a MC move does not result in an energy change. Trial moves are only proposed on the current chain and in the current direction of the sampling. Switching chains and direction happens in separate steps.}
    \label{fig:lifting}
\end{figure*}
Reversible MCMC simulations, parallel or not, operate on a single Markov chain of states on which the ``walker'' can move forward or backward, that is, Monte Carlo trial moves are proposed in either direction. This ``symmetry'' is broken when splitting up the single chain into multiple, lifted chains and proposing trial moves in only one direction, either increasing or decreasing the energy. Fig.~\ref{fig:lifting} shows an example of lifting for a 2D Ising-type model with the Hamiltonian
\begin{equation}
    \mathcal{H}=-\sum_{\left<i,j\right>}J_{ij}\,s_i\,s_j\,,
    \label{eq:Ising}
\end{equation}
where $s_i$ is the individual spin on site $i$ that can have a value of $\pm1$ and $\left<i,j\right>$ are all nearest-neighbor pairs. The interactions $J_{ij}$ can take values of $\pm1$, independently for every pair $\left<i,j\right>$.
Similar to an earlier implementation by Fernandes and Weigel~\cite{weigel2010cpc}, we introduce chains corresponding to the five possible energy changes $\Delta E=\{0,\pm4,\pm8\}$ that can result from a single spin flip. Each chain has a ``direction'', which is either ``forward'' ($\sigma=+1$) or ``backward'' ($\sigma=-1$), depending on spin flips increasing or decreasing the energy, respectively\footnote{Note that we use a slightly different definition of a chain here compared to previous works such as Ref.~\cite{weigel2010cpc}, for ease of presentation. Instead of defining a chain to include both the $\sigma = \pm 1$ directions, we treat each direction as a separate chain.}. We can hence label the chains by ($\sigma,\left|\Delta E\right|$). For each chain, we store the full list of MC moves (i.e., the sites and spin flips) that would result in the designated energy change.

To simulate using such a setup, we also adopt the ``lifting parameter'' $\theta$, the probability to jump between chains after each trial move. Furthermore, we change the direction of the walk ($\sigma\rightarrow -\sigma$) whenever a trial move is rejected; and only then. The recipe for the simulations presented here follows the one that was used in~\cite{weigel2010cpc}, adapted for multicanonical simulations: 
\begin{itemize}
    \item Choose a MC move from the current chain ($\sigma,\left|\Delta E\right|$) and update with the acceptance probability
        \begin{equation*}
            p=\min\left[1,\frac{N^{\textrm{before}}_{\sigma |\Delta E|}}{N^{\textrm{after}}_{-\sigma |\Delta E|}}\,\frac{W(E^{\textrm{after}})}{W(E^{\textrm{before}})}\right]\,.
        \end{equation*}
        ``Before'' and ``after'' refer to the spin flip, that is, $N^{\textrm{before}}_{\sigma |\Delta E|}$ is the number of possible MC moves on the current chain ($\sigma,|\Delta E|$) before the spin is flipped and $N^{\textrm{after}}_{-\sigma |\Delta E|}$ is the number of possible MC moves on the chain in the opposite direction ($-\sigma,|\Delta E|$) after the spin would be flipped. In the same way $E^{\textrm{before}}$ and $E^{\textrm{after}}$ are the energies of the system before and after the spin flip.
    \item If the proposed spin flip was rejected in the previous step, reverse the direction: $\sigma \rightarrow -\sigma$. If there is no MC move available on the current chain and in the current direction, that is $N^{\textrm{before}}_{\sigma |\Delta E|}=0$, the spin flip attempt is automatically ``rejected'' and the direction $\sigma$ is changed.
    \item With a probability $\theta$, randomly jump to a new chain that maintains the current direction~$\sigma$.
\end{itemize}
At any given time step, we only sample from the set of possible MC moves corresponding to the current chain and direction. Therefore we maintain and continuously update a list for each chain containing all spins which would change the energy of the system by the corresponding amount $\Delta E$, if flipped. We create these lists once at the beginning of the simulation. After each spin flip, if necessary, we remove the flipped spin and all its neighbors from their current lists, calculate their local energy changes if they were individually flipped, and reassign them to the corresponding chains. 
The additional bookkeeping necessary for lifting hence consists of the following steps:
\begin{itemize}
    \item After a spin was chosen for a MC trial move, calculate $N^{\textrm{after}}_{-\sigma |\Delta E|}$. To do so, go over that spin and all its neighbors and check if they would be added to the chain ($-\sigma,|\Delta E|$) if the trial move would be accepted\footnote{For the trial spin itself this would always be true. If it was flipped, flipping the spin directly \textit{again} would restore the current configuration.}. If so, increase the counter $N^{\textrm{after}}_{-\sigma |\Delta E|}$ accordingly.
    \item After a trial move is accepted, go over that spin and all its neighbors, delete them from their previous chain and assign them to their new chain. This is, in principle, straightforward as we know the energy change if a spin is flipped, or when a neighbor is flipped. However, these lists are not sorted and they necessarily grow with the total number of spins in the system. In practice, the cost for removing a spin from a list would scale with system size if maintaining sorted lists or using search operations to find a spin in a list. Such costly operations can be avoided though by using an auxiliary lattice to separately store the position of each spin in their list, keeping the amount of bookkeeping computations \textit{per Monte Carlo move} constant. 
\end{itemize}
In summary, even though bookkeeping makes each MC move computationally more expensive compared to reversible sampling, it is not prohibitively complex. We further note that there have been other MC sampling schemes, like the ``$N$-fold way''~\cite{landau2014book}, where similar procedures are necessary.

\section{\label{sec:IsingResults}Results}

We apply the multicanonical lifting-scheme to two Ising-type spin models, both on a 2D-square lattice with nearest neighbor interactions and the Hamiltonian given in Eq.~(\ref{eq:Ising}).
\begin{itemize}
    \item Classical Ising model~\cite{ising25zph}: $J_{ij}=1$ for all nearest neighbor pairs $\left<i,j\right>$.
    \item Edwards--Anderson Ising spin glass~\cite{ea75jph}: $J_{ij}=\pm 1$, randomly chosen for each bond (nearest neighbor pair) with equal probability for $J_{ij}=+1$ and $-1$.
\end{itemize}

\subsection{Comparison to exact solution: the Ising model}

\begin{figure}[t]
    \centering
    \includegraphics[width=\columnwidth]{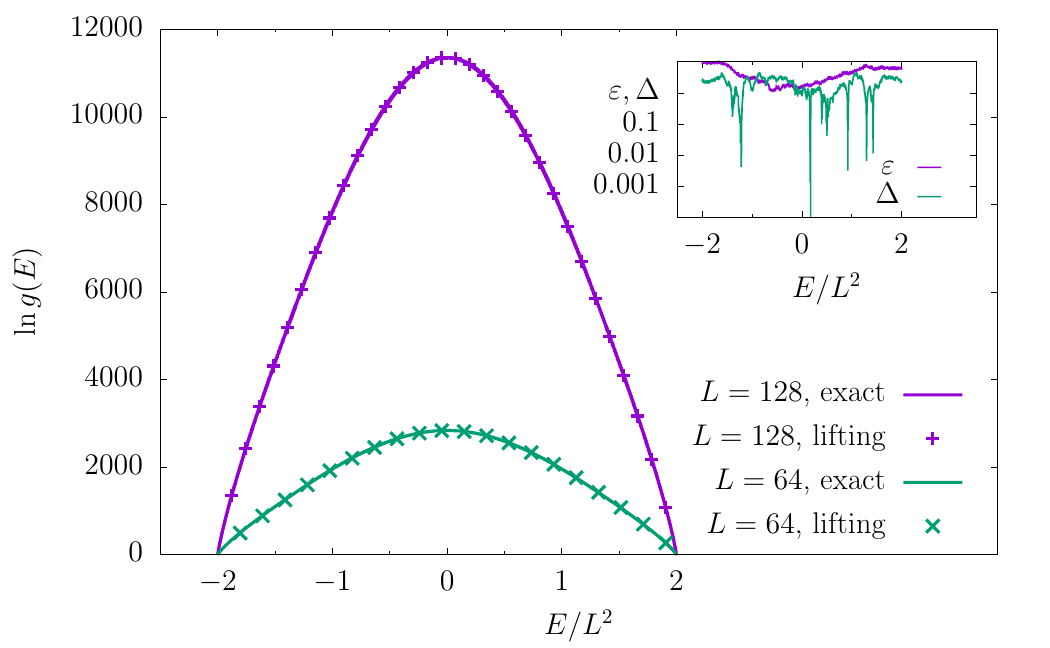}
    \caption{Logarithm of the density of states for $L=64$ and $L=128$ as estimated by lifted multicanonical sampling at $\theta=1$ (solid lines) compared with the known, exact solution (symbols, shown at intervals for clarity). The inset shows the statistical error, $\varepsilon$, from multiple runs and the absolute deviation, $\Delta$, of their average from the exact solution ($\Delta=\lvert\overline{\ln g(E)}-\ln g_{\mathrm{exact}}(E)\rvert$) for $L=128$.}
    \label{fig:gE_comp}
\end{figure}
\begin{figure}[t]
    \centering
    \includegraphics[width=\columnwidth]{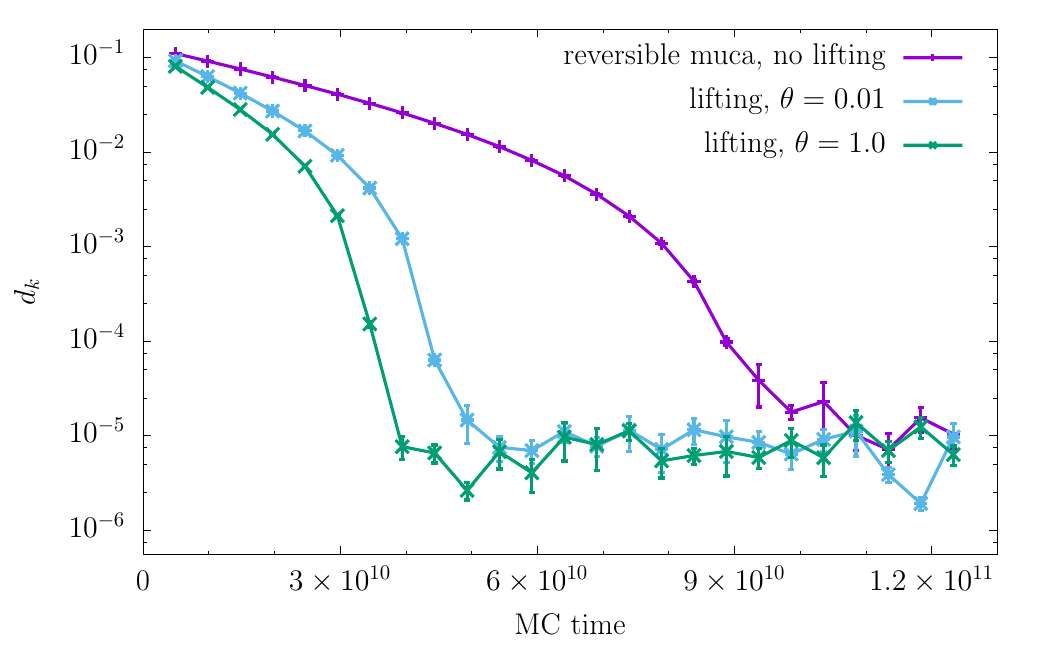}
    \caption{Kullback--Leibler divergence between the true logarithmic density of states of the $L=128$ 2D Ising model and the logarithm of the instantaneous, estimated density of states as a function of Monte-Carlo time. Different curves compare the convergence of ``lifted'' multicanonical simulations at different lifting parameters $\theta$ with the conventional multicanonical sampling without lifting. Error bars were obtained from independent runs for each setting.}
    \label{fig:KL128}
\end{figure}
To verify the correctness of the lifting scheme we apply it to the 2D Ising model with $L\times L$ spins. The logarithm of the density of states calculated from the estimators for the multicanonical weights for $L=64$ and $L=128$ is shown in Fig.~\ref{fig:gE_comp} and compared to the exact solution~\cite{beale96prl}.
To demonstrate the convergence towards the known solution we compute the Kullback--Leibler (KL) divergence~\cite{kullback1951ams}, which has been shown useful for such purposes~\cite{gross2017cpc}:
\begin{equation}
    d_k=\sum_x P(x)\,\textrm{ln}\left[\frac{P(x)}{Q(x)}\right].
\end{equation}
The KL divergence measures the degree of discrepancy between two probability distributions $Q(x)$ and $P(x)$. Since we are particularly interested in the convergence of the exponentially-suppressed regions of the energy space, i.e., the tails of the density of states, we calculate the KL divergence between the logarithm of the exact density of states $g_\textrm{exact}(E)$ and that of the estimated density of states from the multicanonical simulation. The 
estimated density of states is, when converged, the inverse of the simulation weights (see Eq.~\ref{eqn:muca}). Therefore, $P(x)=\mathrm{ln}\,g_\textrm{exact}(E)$, and $Q(x)=-\,\mathrm{ln}\,W(E)$.
Both $P(x)$ and $Q(x)$ are further normalized such that $\sum_x P(x) = \sum_x Q(x) = 1$. Fig.~\ref{fig:KL128} shows the convergence of the density of states estimated from independent multicanonical simulations, reversible and non-reversible (lifted), towards the exact solution as measured by the KL divergence for the $L=128$ Ising model. The MC time is measured in single-spin flips. Aside from the non-reversible runs converging to the true result in the same way as a conventional, reversible simulation and reaching the same, final accuracy, they can do so much faster; from Fig.~\ref{fig:KL128} we estimate a speedup of about three, in our case. We will discuss this in more detail below.

\subsection{Accelerating the exploration of the complete energy range and ground-state search}

\begin{figure}[b!]
    \centering
    \includegraphics[width=\columnwidth]{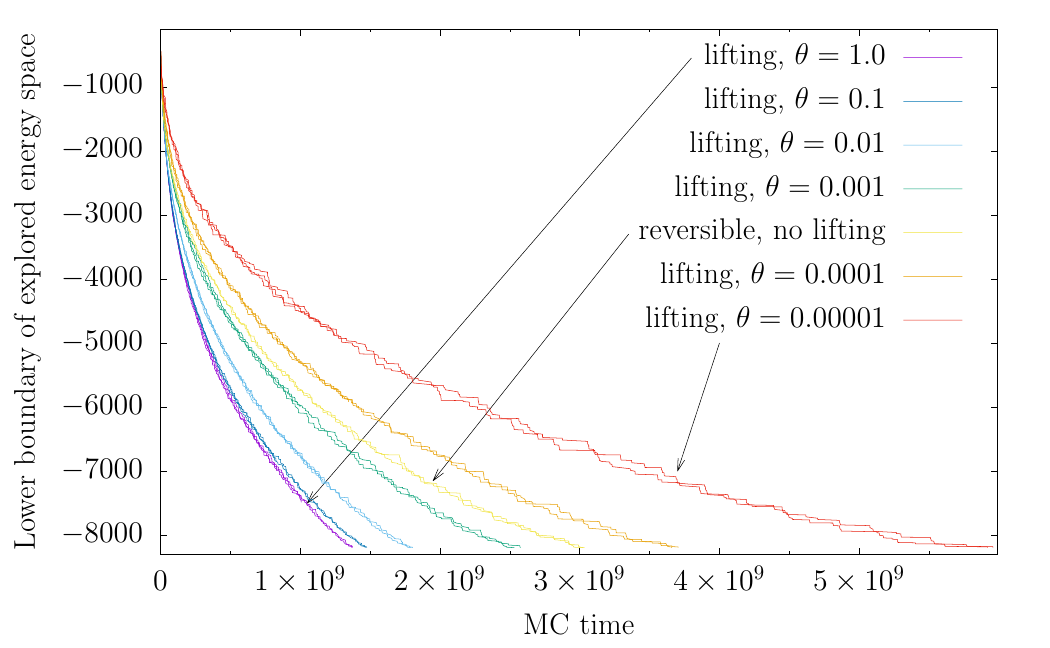}
    \includegraphics[width=\columnwidth]{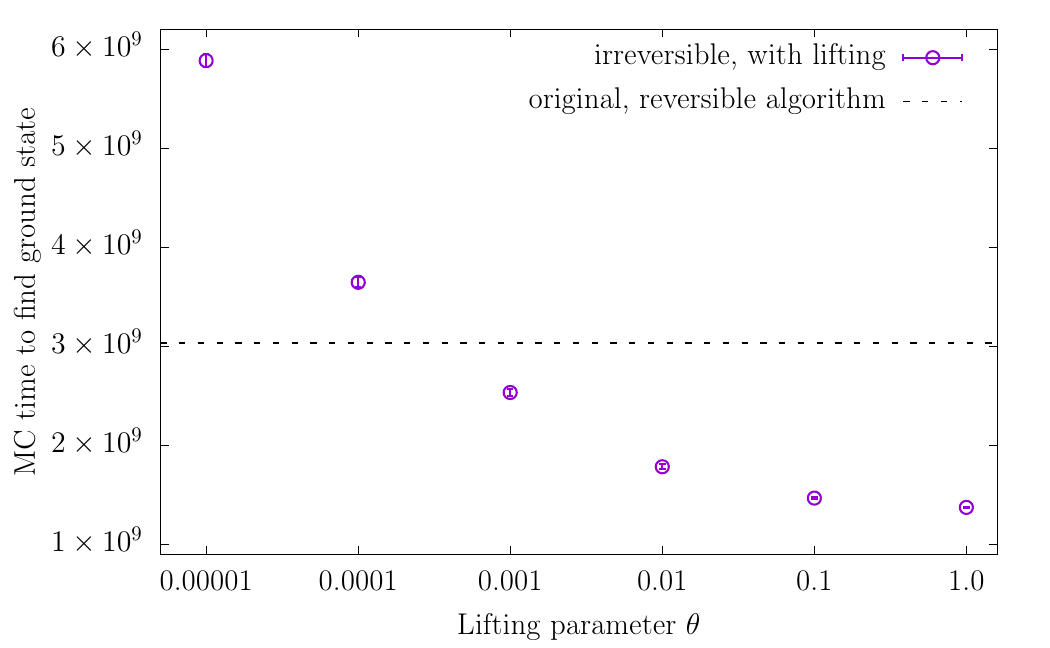}
    \caption{Top: Time evolution of the lowest energy found for the $L=64$ Ising model at different lifting parameters, showing how the explored energy range widens over time. The upper boundary of the explored energy range (at positive energies) is symmetric to the lower boundary and not shown. Three individual runs are shown for each value of $\theta$ to show the consistency of the method, the curves end when the ground state is found for the first time. Bottom: The average Monte-Carlo (MC) time needed to find the ground state for the first time depending on the lifting parameter. The horizontal dashed line denotes the MC time needed for the original multicanonical algorithm to find the ground state for the first time.\vspace{-.6mm}}
    \label{fig:GS}
\end{figure}
Before data can be collected from flat-histogram sampling methods, typically referred to as the ``production run'', the whole range of, in our case, energies has to be explored. For systems with a known ground-state energy, this means the ground state must be found. Similar to other methods like Wang--Landau sampling~\cite{wang2001prl:wl,wang2001pre:wl}, a~multicanonical recursion starts with an initial configuration at an arbitrary energy. During the initial phase in which the simulation weights are determined, the boundaries of the explored energy range will continuously expand. In Fig.~\ref{fig:GS}\,(top) we show the low-energy boundary of the explored energy range as it shifts over time for various values of lifting parameter $\theta$ for the $L=64$ Ising model. In other words, we show the lowest known energy visited at any given time.\footnote{We discuss only the lower bound of the energy range here. The broadening of the energy range is symmetric about 0 for the Ising model when starting from a random initial configuration. We confirmed that the upper boundary behaves identically within statistical fluctuations, but do not show this data since it does not provide any additional \hbox{information}.} The lower end of the plot marks the known ground-state energy $E_{\mathrm{GS}}=-8192$ for this system. In Fig.~\ref{fig:GS}\,(bottom) we plot the time it took to find the ground state for the first time, again depending on the lifting parameter $\theta$. While a significant speedup in exploration of the energy space is seen when using a suitable lifting parameter, a non-reversible scheme could also perform worse than a~reversible simulation with a poor choice of a lifting parameter, that is, if jumps between chains do not happen frequently enough.

\begin{figure}[b!]
    \centering
    \includegraphics[width=\columnwidth]{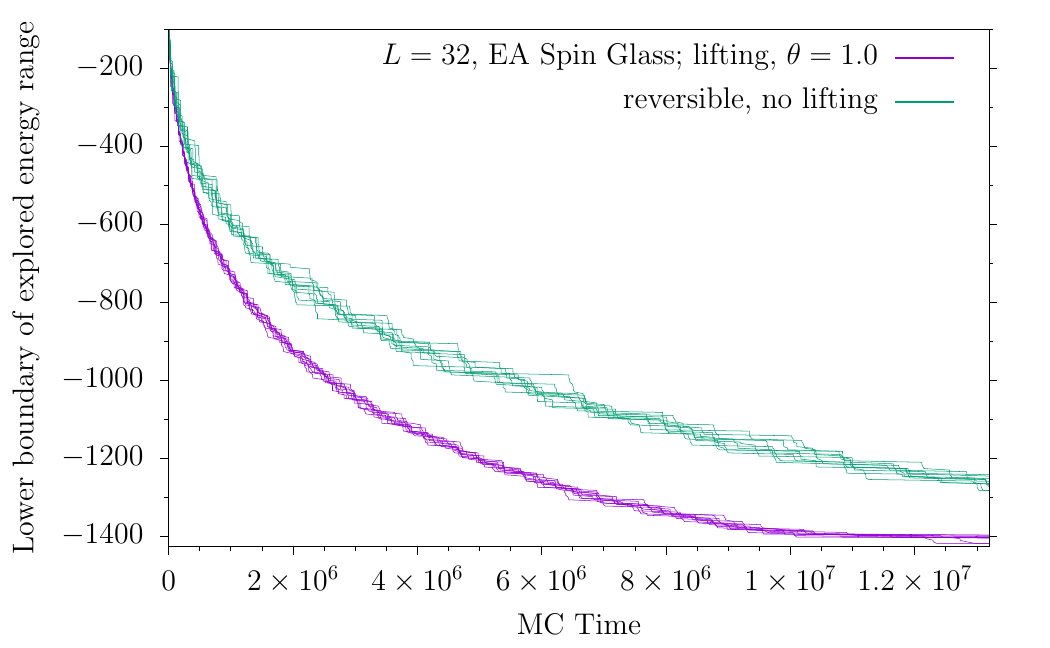}
    \includegraphics[width=\columnwidth]{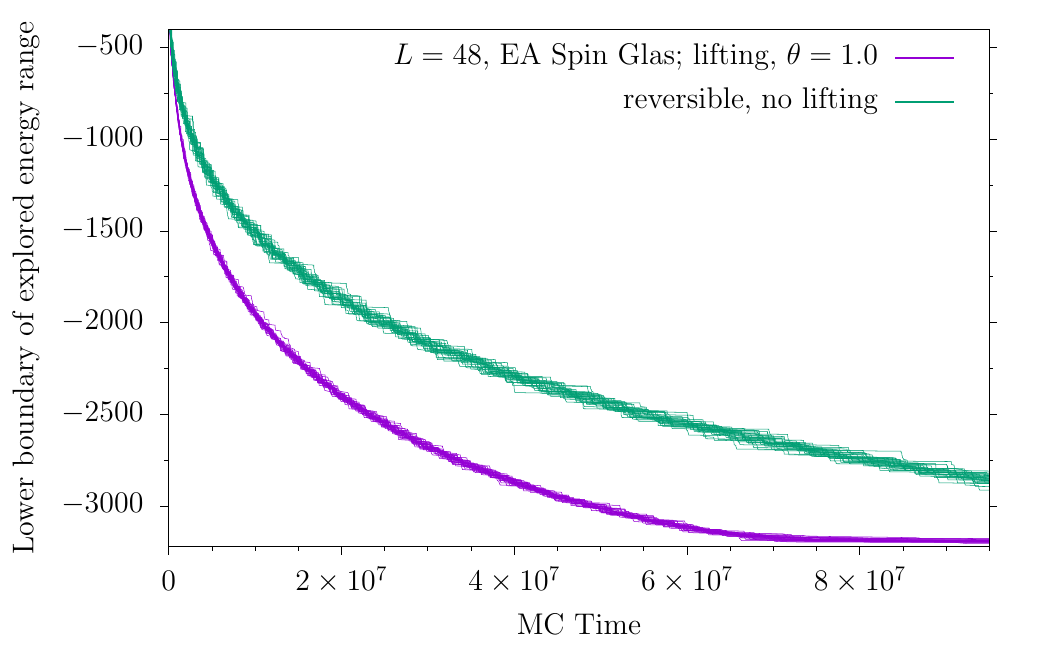}
    \caption{Top: Time evolution of the lowest energy found for an instance of the $L=32$ Edwards--Anderson spin glass model, showing how the explored energy range widens over time. To demonstrate the consistency of the method, multiple individual runs are shown for both the conventional, reversible and the non-reversible, lifted multicanonical scheme at $\theta=1.0$. The lower boundary of the plot area corresponds to the ground-state energy for this spin-glass instance. Bottom: The corresponding plot for an $L=48$ spin-glass \hbox{instance}.\vspace{-1.4mm}}
    \label{fig:time_evol_EA}
\end{figure}
\begin{figure}[t!]
    \centering
    \includegraphics[width=\columnwidth]{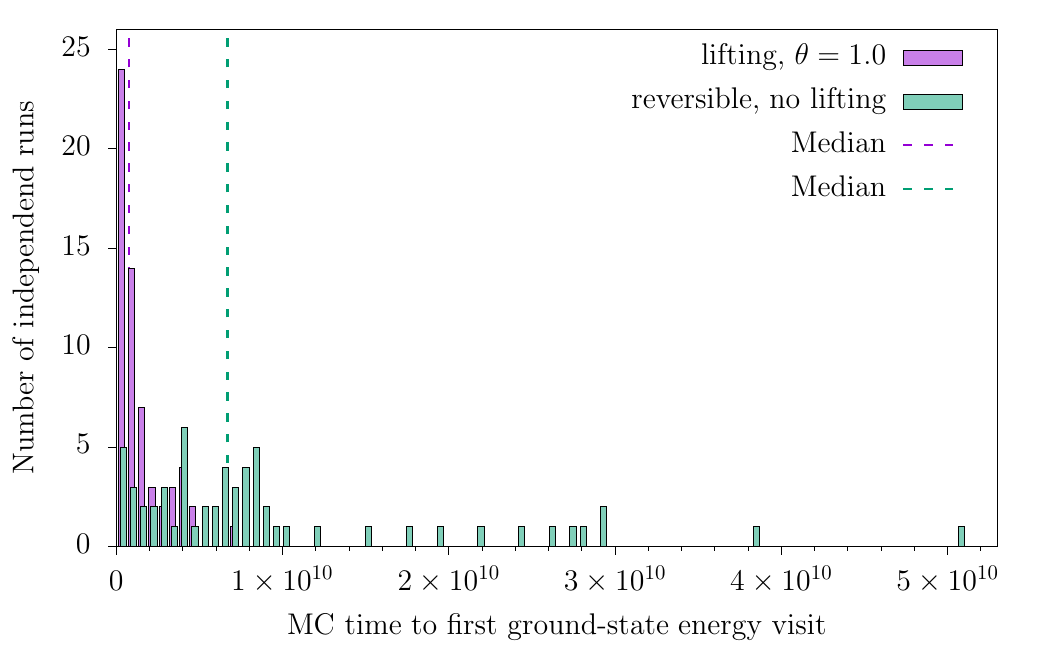}
    \includegraphics[width=\columnwidth]{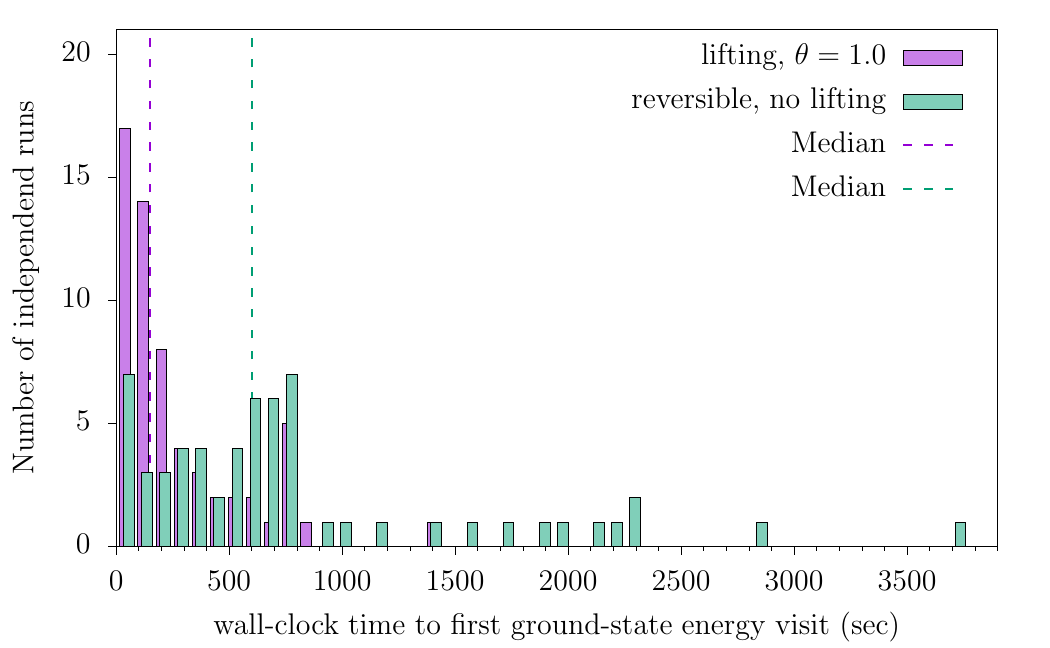}
    \caption{Top: Histograms of MC times to find the ground-state energy of an instance of the $L=32$ Edwards--Anderson spin glass model for the first time using conventional multicanonical sampling and the lifted algorithm. We ran 60 independent simulations each, with different, random initial spin configurations. The median time (vertical, dashed lines), that is, the time after which 30 of the runs found the ground-state energy, is almost one order of magnitude smaller for the lifted implementation. Besides the shorter median time, note the larger spread and outliers in the large-time tails of the histograms for conventional multicanonical sampling. Bottom: The corresponding histogram of real, wall-clock times. With the additional computational overhead for the lifting, the median time to find the ground state is still smaller by a factor of 4. The ground-state energies were independently confirmed using the McGroundstate solution server for spin glasses~\cite{CJMM22_server,CJMM22}.}
    \label{fig:GS_EA}
\end{figure}

In Figs.~\ref{fig:time_evol_EA} and \ref{fig:GS_EA} we show a similar analysis for two instances of an Edwards--Anderson (EA) Ising spin glass with linear sizes $L=32$ and $L=48$. For the EA model, ground-state energies can also be computed exactly~\cite{hartmann2011jsp,CJMM22}. For the instances used here the ground-state energies are $E_{\mathrm{GS}}^{\mathrm{EA}}=-1426$ and $E_{\mathrm{GS}}^{\mathrm{EA}}=-3222$, respectively, which were independently confirmed using~\cite{CJMM22_server}. Similar to the runs for the Ising model shown in Fig.~\ref{fig:GS}\,(top), we find significant speedups in exploring the whole energy range, see Fig.~\ref{fig:time_evol_EA}. In Fig.~\ref{fig:GS_EA} we show histograms of the times needed to reach the ground-state energy of the $L=32$ instance for 60 independent runs. The median time to find a ground state, measured in Monte Carlo time, with lifting is about one order of magnitude smaller compared to conventional runs without lifting (Fig.~\ref{fig:GS_EA}\, top: $8.0\times 10^8$ vs. $6.7\times 10^9$; vertical, dashed lines). This is the most significant individual speedup we found in this study and could be very relevant for applications that search for ground states in similar quadratic unconstrained binary optimization (QUBO) \hbox{problems}~\cite{kochenberger2014a}. To take the additional computational overhead for lifting into account, we also measured the real, wall-clock time to find a ground state. The median of this time is still 4 times smaller compared to a conventional, multicanonical run without lifting (Fig.~\ref{fig:GS_EA}\, bottom: $151$\,s vs. $601$\,s; vertical, dashed lines). That is, a MC move is about twice as expensive with the bookkeeping needed for the lifting procedure in our implementation\footnote{The measurements of real time (or CPU time) of course strongly depend on the actual implementation of the spin lists and the operations on them. We are not presenting a performance optimization study here, just a proof-of-concept. We are convinced that the computing overhead for the lifting bookkeeping can be reduced even more with further code optimization.}. However, as discussed above, the overhead per spin flip, and consequently increase in real time per MC step, is of order $\mathcal{O}(1)$; that is, it does not increase with system size.

\subsection{Round-trip times after finding the ground state}

\begin{figure}[b!]
    \centering
    \includegraphics[width=\columnwidth]{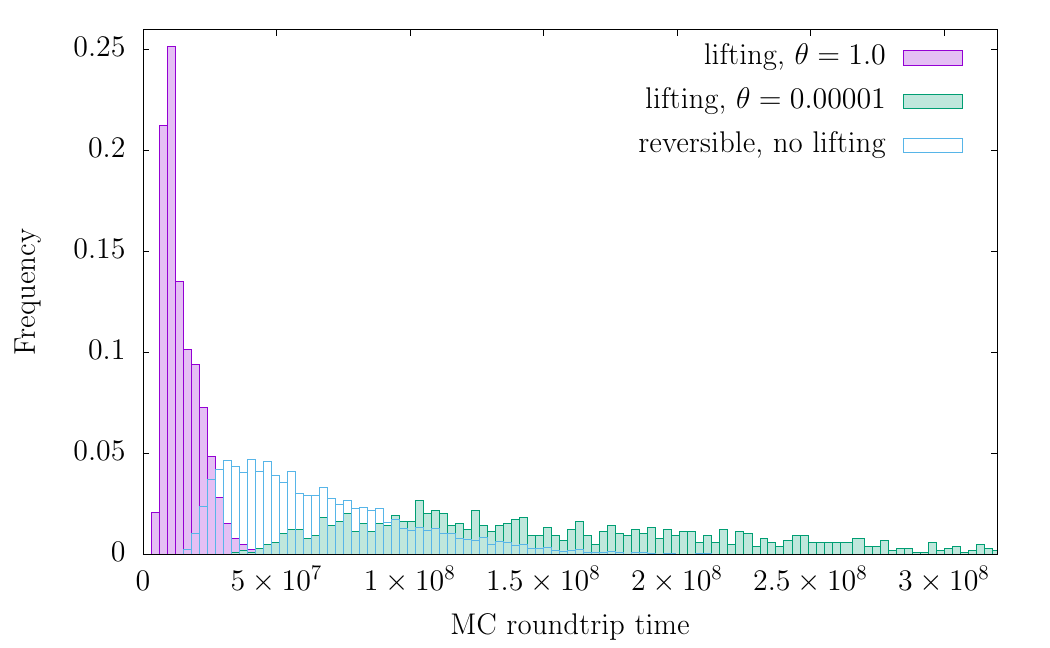}
    \includegraphics[width=\columnwidth]{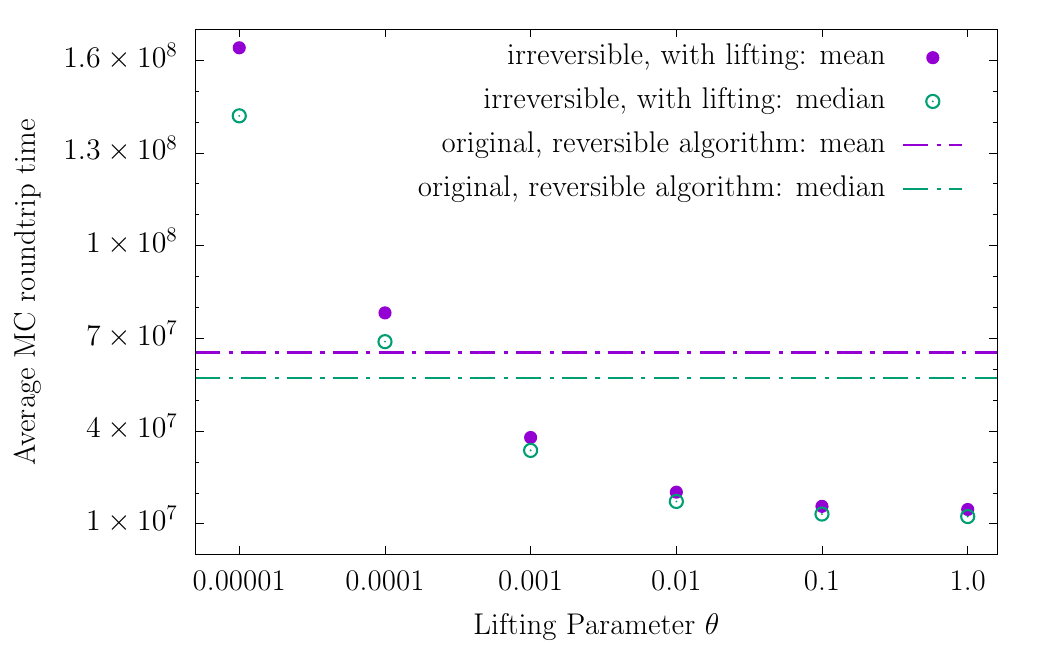}
    \caption{Top: Histograms of round-trip times for the 2D Ising model after the ground-state has been found. Bottom: The average and median round-trip times for different lifting parameters $\theta$.\vspace{-1.4mm}}
    \label{fig:roundtrip}
\end{figure}
\begin{figure*}[t]
    \centering
    \includegraphics[width=\columnwidth]{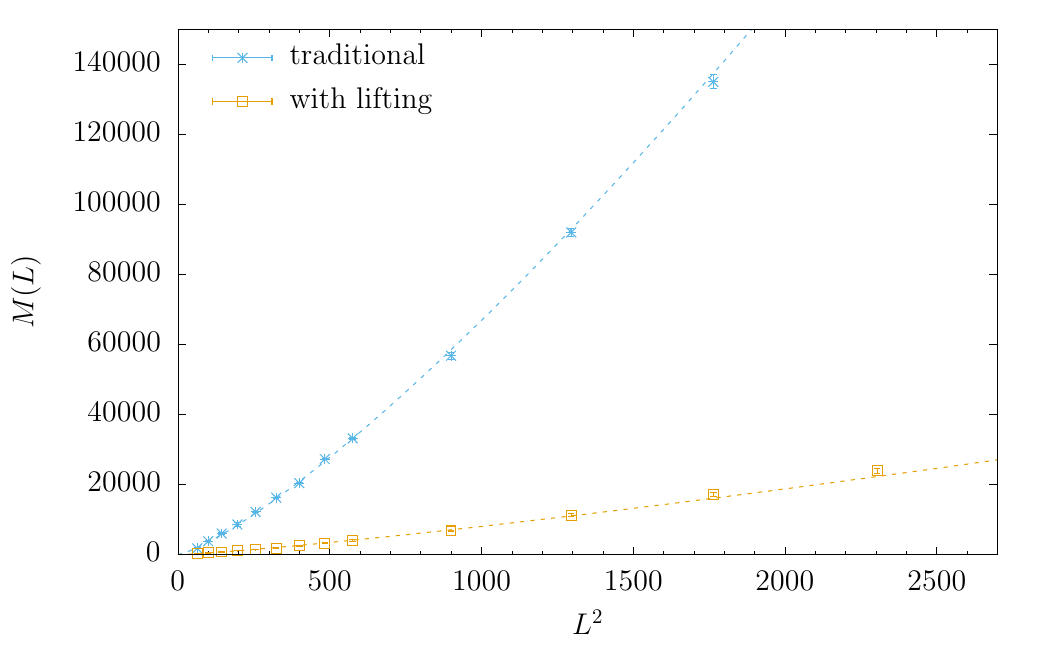}
    \includegraphics[width=\columnwidth]{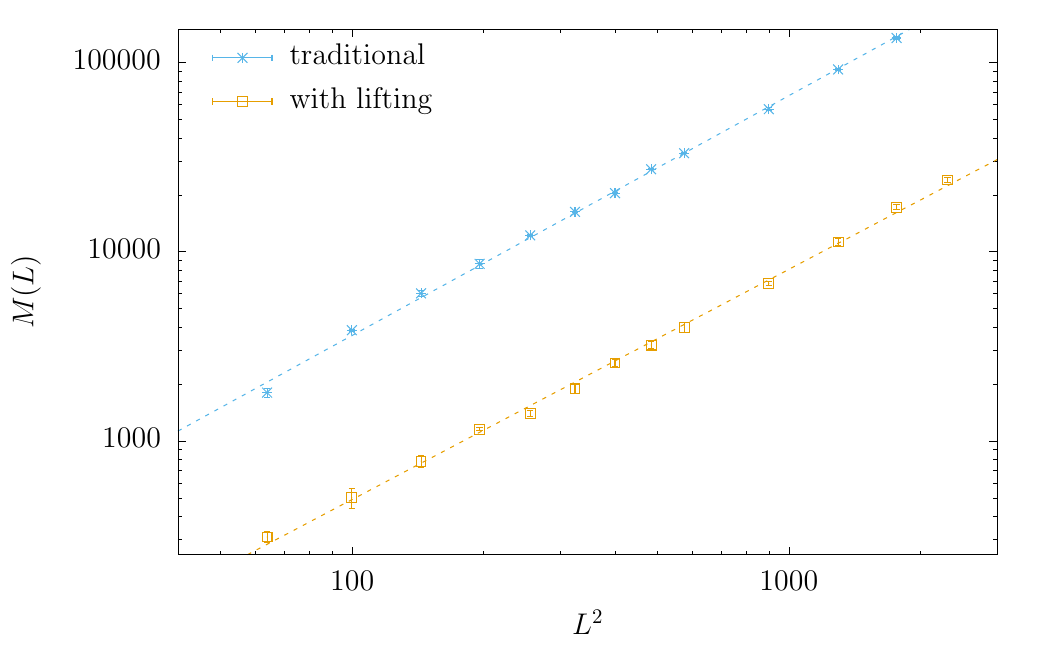}
    \caption{Left: The optimal number of sweeps per iteration, $M(L)$, as a function of the system size for the 2D Ising model. Right: The same scaling on a logarithmic scale, showing the power-law behavior. The exponent measured for traditional multicanonical runs agrees with previous results~\cite{zierenberg2013cpc}. }
    \label{fig:muca_param}
\end{figure*}
In order to evaluate the sampling performance after the ground state has been found, we measure the round-trip time -- the average time it takes for the random walker to complete a walk over the entire energy range and return to the starting energy. Fig.~\ref{fig:roundtrip}\,(top) shows the histograms of round-trip times for the 2D Ising model at two values of the lifting parameter~$\theta$, compared to a reversible simulation. Fig.~\ref{fig:roundtrip}\,(bottom) shows the average and median round-trip times calculated from those histograms for different values of~$\theta$. As one might expect, the speedup is very similar to the one observed for the ground-state search (Fig.~\ref{fig:GS}, bottom). Aside from the speedup, we also note that the variation of the round-trip times is dependent on the lifting parameter~$\theta$, the histogram for $\theta=1.0$ is much sharper compared to $\theta=0.00001$, meaning the run time is more predictable compared to traditional MC simulations. 

\subsection{The optimal number of sweeps}

As discussed in Sec.~\ref{sec:method}, sampling weights for multicanonical simulations can be obtained in different ways; a common method is to calculate them iteratively~\cite{berg1996jsp,janke1998pa,berg2003cpc}. Within each iteration the simulation weights are held constant and a histogram of the energy (or other order parameters of interest~\cite{berg1993prb,janke1998pa,nussbaumer2008pre,millar2023ress}) is recorded. The weights are then updated after a given time $\tau$, before entering the next iteration step. Ideally, this time $\tau$ is chosen such that the whole, previously explored range of the order parameter is sampled adequately. It is therefore related to the autocorrelation time and the average acceptance rate of the multicanonical update. In principle, this would be captured by a function of the width of the currently covered energy range; see~\cite{zierenberg2013cpc} for further discussions. To directly compare results from our study with previous work, we exactly follow the heuristic method presented in~\cite{zierenberg2013cpc}, which aims at reasonably estimating this behavior: we measure the optimal number of sweeps per iteration, $M_{\textrm{opt}}(L)$, so that the mean number of iterations until convergence to a flat histogram is minimal. Fig.~\ref{fig:muca_param} shows the scaling of $M_\textrm{opt}$ with system size~$L^2$ for the 2D Ising model, from which we obtain the following scaling relations:
\begin{align}
    M_{\textrm{opt}}^{\textrm{trad}}(L)&=10.4(1.0)\times L^{2.54(3)}\label{eq:Mtrad},\\
    M_{\textrm{opt}}^{\textrm{lift}}(L)&=1.8(0.2)\times L^{2.44(3)}\label{eq:Mlift}.
\end{align}
The power law behavior is expected to have the form $L^{D+x}$ and our measured exponent for a traditional multicanonical run without lifting ($M_{\textrm{opt}}^{\textrm{trad}}(L)$; Eq.~\ref{eq:Mtrad}) agrees with previous findings~\cite{zierenberg2013cpc}.\footnote{The prefactor might depend on the specific implementation of the multicanonical update. Our values might differ because the weight update in between iterations in~\cite{zierenberg2013cpc} included some extrapolation at the boundaries of the explored energy range; which we did not employ in our simulations.} For simulations with lifting we find a significant difference in the prefactor and also a potentially somewhat lower exponent, see Eq.~(\ref{eq:Mlift}). We found a speed-up of larger than six in the overall convergence time (in Monte Carlo time; independent of system size) by measuring the time needed to converge to a flat histogram from the initial canonical distribution and setting the $\tau$ parameter to its optimal value for both lifted and non-lifted simulation. The main contributor to this overall speedup is the smaller time $\tau$ needed to proceed to the next iteration and minimize the overall convergence time for lifted \hbox{simulations}.

\section{Summary}

Traditional flat-histogram sampling methods often show a slowing down in convergence at the later stage of a simulation due to random diffusion in the order parameter space. Non-reversible simulations are designed to reduce this behavior by avoiding such unrestricted random walks and hence can significantly speed up Monte Carlo (MC) sampling. Performance improvements (in MC time) depend on the physical system and the measured observable. For example, we find speed-ups of a factor of about six for the overall convergence to a flat histogram in the 2D Ising model and up to a factor of ten for finding lowest energies in 2D Edwards--Anderson spin glasses.

We employed a modified lifting scheme and found that it was most beneficial to jump between lifted chains frequently while maintaining the direction of the sampling (towards higher or lower energies, in our case), until a proposed MC move is rejected. Including or excluding microcanonical spin flips, i.e., the $\Delta E=0$ chain, does not affect the general conclusions of this work. We did not observe any differences in the accuracy of the results or the performance, whether or not microcanonical moves were included. If ergodicity breaking is a concern, however, such moves should be included. We ran simulations with and without microcanonical updates and obtain the correct $g(E)$ for the Ising model and consistently find ground states of EA spin glass instances both ways. That would not be the case if configurations would be excluded from the \hbox{sampling}.

It has been noted before~\cite{weigel2010cpc}, and we confirm, that the bookkeeping necessary for this type of simulation is considerable. However, this can be implemented in a way that the overhead cost per MC move is constant, that is, it does not increase with system size, and the speedup gained by reducing the diffusive behavior of a random walk in the parameter space can still outweigh this additional cost. In our implementation, a MC move in the lifting scheme was about twice as expensive in real time, compared to a non-lifted simulation. Furthermore, a lifting scheme for spin systems like the ones used here can be readily implemented in a massively parallel implementation of multicanonical sampling as described in~\cite{gross2017cpc}.

\begin{acknowledgments}
We thank A. Hartmann for valuable discussions during this project. This work was supported by the Los Alamos National Laboratory (LANL) Laboratory Directed Research and Development (LDRD) Program. LANL is managed by Triad National Security, LLC for the U.S. DOE’s NNSA, under contract 89233218CNA000001. This research used resources provided by the Darwin testbed at LANL which is funded by the Computational Systems and Software Environments subprogram of LANL's Advanced Simulation and Computing program (NNSA/DOE). Some data was produced on the University of North Georgia's Pando computing\break{} cluster.
\end{acknowledgments}

\bibliography{main}

\end{document}